\documentclass[a4paper]{article}

\usepackage{INTERSPEECH2020}

\usepackage{hyperref}

\usepackage{tikz}
\usepackage{multirow}

\usetikzlibrary{calc,trees,positioning,arrows,chains,shapes.geometric,%
    decorations.pathreplacing,decorations.pathmorphing,shapes,%
    matrix,shapes.symbols}

\tikzset{%
  block/.style  = {draw, thick, rectangle, minimum height = 3em, minimum width = 3em},
  neuron/.style = {draw, circle, minimum height=2em},
  sum/.style    = {draw, circle}, 
  layer/.style  = {rectangle, draw=black, minimum height=1.7em, minimum width=20mm, text centered},
}

\usepackage{amsmath,amssymb}
\DeclareMathOperator{\adain}{AdaIN}

\newcommand{\vect}[1]{\mathbf{#1}}

\newcommand{\matr}[1]{\mathbf{#1}}

\newcommand{\vc}[0]{\vect{c}}

\newcommand{\vm}[0]{\vect{m}}

\newcommand{\vw}[0]{\vect{w}}

\newcommand{\vx}[0]{\vect{x}}
\newcommand{\vy}[0]{\vect{y}}
\newcommand{\vz}[0]{\vect{z}}

\newcommand{\mC}[0]{\vect{C}}

\title{Conditional Spoken Digit Generation with StyleGAN}
\name{Kasperi Palkama, Lauri Juvela, Alexander Ilin}
\address{
  Aalto University, Espoo, Finland
}
\email{kasperi.palkama@gmail.com, lauri.juvela@aalto.fi, alexander.ilin@aalto.fi}

\begin{document}

\maketitle
\begin{abstract}
This paper adapts a StyleGAN model for speech generation with minimal or no conditioning on text. 
StyleGAN is a multi-scale convolutional GAN capable of hierarchically capturing data structure and latent variation on multiple spatial (or temporal) levels. 
The model has previously achieved impressive results on facial image generation, and it is appealing to audio applications due to similar multi-level structures present in the data. 
In this paper, we train a StyleGAN to generate mel-spectrograms on the Speech Commands dataset, which contains spoken digits uttered by multiple speakers in varying acoustic conditions.
In a conditional setting our model is conditioned on the digit identity, while learning the remaining data variation remains an unsupervised task. 
We compare our model to the current unsupervised state-of-the-art speech synthesis GAN architecture, the WaveGAN, and show that the proposed model outperforms
according to numerical measures and subjective evaluation by listening tests.
\end{abstract}


\section{Introduction}

\label{sec:intro}

Speech synthesis using neural networks has seen rapid advancements in recent years, and deep learning is a fundamental component for building state-of-the-art text-to-speech (TTS) applications \cite{tacotron2}. A major factor in these improvements has been the adoption of generative models, such as WaveNet \cite{oord2016-wavenet-arxiv}. Conversely, speech data is interesting for benchmarking the performance of generative models, as it contains both deterministic structure and stochastic variation at multiple levels, ranging from utterance, word and phoneme level all the way to short-time segmental signal characteristics. 

Generative adversarial networks (GANs) \cite{gan} have attracted enormous research attention since their introduction and have since achieved high synthesis quality in the image domain \cite{progan, largescalegan}. In the audio and speech domain, research on GANs for unconditional (or weakly conditioned) synthesis tasks has been fairly limited. 
Convolutional GAN has demonstrated a capability to synthesize various raw audio sounds (WaveGAN) \cite{wavegan}, 
although at limited quality, while \cite{gansynth} applied GANs on pitch-conditioned instrument sound synthesis. 

Meanwhile in text-to-speech synthesis, GANs have been applied to the two sub-problems that constitute the current state-of-the-art: first, an acoustic model learns a mapping from a text sequence to acoustic features (i.e., mel-spectrogram in \cite{tacotron2}), and second, a waveform generator model maps the acoustic features to a speech waveform. For the acoustic modeling task, adversarial training has been used as an auxiliary objective alongside conventional regression \cite{Saito2018-gan-tts, zhao2018-wgan-speech-synthesis} or as a generative post-filter to add stochasticity to regression-based predictions \cite{Kaneko2017-gan-postfilter}.
GAN-based waveform synthesis models that are capable of fast parallel inference have recently been proposed as alternative to autoregressive WaveNets for the latter task \cite{Juvela2019-gelp, kumar2019-melgan, binkowski2020-hifi-tts-gans}. However, all these approaches use strong conditioning that is either trivially aligned to the generated data or relies on external systems for alignment. 
Furthermore, they typically combine adversarial training with a regression task, which makes it difficult to assess the GAN performance in isolation.

In this work, we propose a GAN-based approach for generating mel-spectrograms of spoken digits from the speech commands dataset \cite{sc09} using only a global conditioning on the digit identity and a purely adversarial training objective. In addition to the labeled variation of digits, the data contains a large amount of unlabelled variation in terms of different speakers and acoustic environments. This makes the dataset interesting for evaluating GANs ability of capturing the variation in an unsupervised manner. On the other hand, the task is connected to TTS acoustic modeling, and the present research takes a first step toward building a purely GAN-based acoustic model. The appeal of GAN for acoustic modeling includes not only parallel inference, but also the potential to capture and recreate non-annotated variation in the data.  

The proposed architecture is a conditional version of the style-based generator architecture for generative adversarial networks (StyleGAN) \cite{stylegan}. Experimental results show that the proposed method outperforms a DCGAN baseline in various objective metrics, as well as subjective naturalness evaluation by listening tests.

\section{Model architecture}
\label{sec:model}

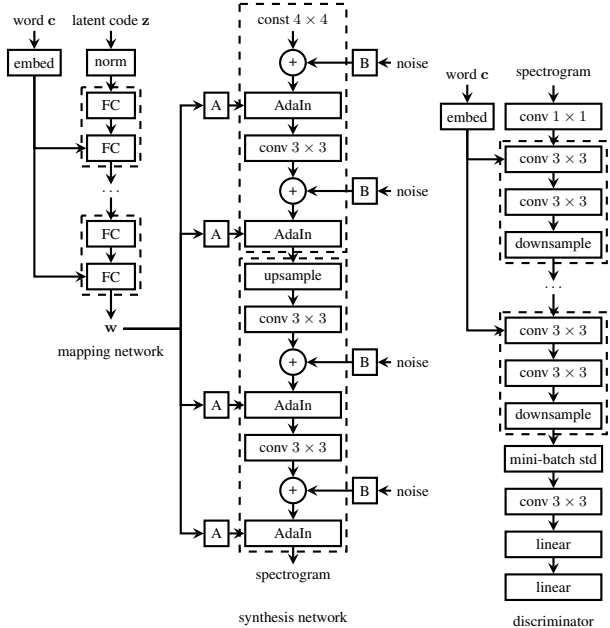
\begin{figure}[t]
\centering
\begin{tikzpicture}[node distance=9mm, thick, scale=0.63, every node/.style={scale=0.63}]
\small
\draw
node at (0, 0) (z) {latent code $\vz$}
node [layer, below of=z, minimum width=10mm] (pn) {norm}

node [layer, below of=pn, minimum width=10mm] (fc1) {FC}
node [layer, below of=fc1, minimum width=10mm] (fc2) {FC}

node [below of=fc2] (dots) {$\ldots$}

node [layer, below of=dots, minimum width=10mm] (fc3) {FC}
node [layer, below of=fc3, minimum width=10mm] (fc4) {FC}

node [below of=fc4, node distance=11mm] (w) {$\vw$}

node [left of=z, node distance=16mm](digit) {word $\vc$}
node [layer, below of=digit, minimum width=11mm] (emb) {embed}

node [below of=w, node distance=5mm] {mapping network}

;

\draw [draw=black, dashed] ($(fc1.north west) + (-1mm, 1mm)$)
      rectangle ($(fc2.south east) + (1mm, -1mm)$);

\draw [draw=black, dashed] ($(fc3.north west) + (-1mm, 1mm)$)
      rectangle ($(fc4.south east) + (1mm, -1mm)$);

\draw [->,>=stealth](z) -- node {} (pn);
\draw [->,>=stealth](pn) -- node {} (fc1);
\draw [->,>=stealth](fc1) -- node {} (fc2);
\draw [->,>=stealth](fc2) -- node {} (dots);
\draw [->,>=stealth](dots) -- node {} (fc3);
\draw [->,>=stealth](fc3) -- node {} (fc4);
\draw [->,>=stealth](fc4) -- node {} (w);

\draw [->,>=stealth](digit) -- node {} (emb);
\draw [->,>=stealth](emb) |- node {} (fc2);
\draw [->,>=stealth](emb) |- node {} (fc4);

\draw
node at (3.8, 0) [minimum width=20mm](const) {const $4 \times 4$}
node [sum, below of=const] (sum1) {+}
node [layer, below of=sum1] (ada1) {AdaIn}
node [layer, below of=ada1] (conv1) {conv $3 \times 3$}
node [sum, below of=conv1] (sum2) {+}
node [layer, below of=sum2] (ada2) {AdaIn}

node [layer, below of=ada2](up1) {upsample}
node [layer, below of=up1] (conv2) {conv $3 \times 3$}
node [sum, below of=conv2] (sum3) {+}
node [layer, below of=sum3] (ada3) {AdaIn}
node [layer, below of=ada3] (conv3) {conv $3 \times 3$}
node [sum, below of=conv3] (sum4) {+}
node [layer, below of=sum4] (ada4) {AdaIn}

node [layer, right of=sum1, minimum width=5mm, node distance=15mm] (b1) {B}
node [layer, right of=sum2, minimum width=5mm, node distance=15mm] (b2) {B}
node [layer, right of=sum3, minimum width=5mm, node distance=15mm] (b3) {B}
node [layer, right of=sum4, minimum width=5mm, node distance=15mm] (b4) {B}

node [right of=b1, node distance=10mm] (n1) {noise}
node [right of=b2, node distance=10mm] (n2) {noise}
node [right of=b3, node distance=10mm] (n3) {noise}
node [right of=b4, node distance=10mm] (n4) {noise}

node [layer, left of=ada1, minimum width=5mm, node distance=16mm] (a1) {A}
node [layer, left of=ada2, minimum width=5mm, node distance=16mm] (a2) {A}
node [layer, left of=ada3, minimum width=5mm, node distance=16mm] (a3) {A}
node [layer, left of=ada4, minimum width=5mm, node distance=16mm] (a4) {A}

node [left of=const, node distance=20mm, inner sep=0](wgen) {}

node [below of=ada4](image) {spectrogram}
node [below of=image] {synthesis network}
;

\draw [draw=black, dashed] ($(const.north west) + (-1mm, 1mm)$)
     rectangle ($(ada2.south east) + (1mm, -1mm)$);

\draw [draw=black, dashed] ($(up1.north west) + (-1mm, 1mm)$)
     rectangle ($(ada4.south east) + (1mm, -1mm)$);

\draw [->,>=stealth](const) -- node {} (sum1);
\draw [->,>=stealth](sum1) -- node {} (ada1);
\draw [->,>=stealth](ada1) -- node {} (conv1);
\draw [->,>=stealth](conv1) -- node {} (sum2);
\draw [->,>=stealth](sum2) -- node {} (ada2);

\draw [->,>=stealth](ada2) -- node {} (up1);
\draw [->,>=stealth](up1) -- node {} (conv2);
\draw [->,>=stealth](conv2) -- node {} (sum3);
\draw [->,>=stealth](sum3) -- node {} (ada3);
\draw [->,>=stealth](ada3) -- node {} (conv3);
\draw [->,>=stealth](conv3) -- node {} (sum4);
\draw [->,>=stealth](sum4) -- node {} (ada4);
\draw [->,>=stealth](ada4) -- node {} (image);

\draw [->,>=stealth](b1) -- node {} (sum1);
\draw [->,>=stealth](b2) -- node {} (sum2);
\draw [->,>=stealth](b3) -- node {} (sum3);
\draw [->,>=stealth](b4) -- node {} (sum4);

\draw [->,>=stealth](n1) -- node {} (b1);
\draw [->,>=stealth](n2) -- node {} (b2);
\draw [->,>=stealth](n3) -- node {} (b3);
\draw [->,>=stealth](n4) -- node {} (b4);

\draw [->,>=stealth](a1) -- node {} (ada1);
\draw [->,>=stealth](a2) -- node {} (ada2);
\draw [->,>=stealth](a3) -- node {} (ada3);
\draw [->,>=stealth](a4) -- node {} (ada4);


\draw [->,>=stealth](w.east) -- ++(1.2,0) |- node {} (a1);
\draw [->,>=stealth](w.east) -- ++(1.2,0) |- node {} (a2);
\draw [->,>=stealth](w.east) -- ++(1.2,0) |- node {} (a3);
\draw [->,>=stealth](w.east) -- ++(1.2,0) |- node {} (a4);

\end{tikzpicture}
\begin{tikzpicture}[node distance=9mm, thick, scale=0.63, every node/.style={scale=0.63}]
\small
\draw
node at (0, 0) (spectrogram) {spectrogram}
node [layer, below of=spectrogram] (conv1) {conv $1 \times 1$}

node [left of=spectrogram, node distance=18mm](digit) {word $\vc$}
node [layer, below of=digit, minimum width=11mm] (emb) {embed}

node [layer, below of=conv1] (conv2) {conv $3 \times 3$}
node [layer, below of=conv2] (conv3) {conv $3 \times 3$}
node [layer, below of=conv3] (ds) {downsample}

node [below of=ds](dots) {$\ldots$}

node [layer, below of=dots] (conv4) {conv $3 \times 3$}
node [layer, below of=conv4] (conv5) {conv $3 \times 3$}
node [layer, below of=conv5] (ds2) {downsample}

node [layer, below of=ds2] (mb) {mini-batch std}
node [layer, below of=mb] (conv6) {conv $3 \times 3$}
node [layer, below of=conv6] (lin1) {linear}
node [layer, below of=lin1] (lin2) {linear}

node [below of=lin2, node distance=7mm] {discriminator}

;

\draw [draw=black, dashed] ($(conv2.north west) + (-1mm, 1mm)$)
      rectangle ($(ds.south east) + (1mm, -1mm)$);

\draw [draw=black, dashed] ($(conv4.north west) + (-1mm, 1mm)$)
      rectangle ($(ds2.south east) + (1mm, -1mm)$);

\draw [->,>=stealth](spectrogram) -- node {} (conv1);
\draw [->,>=stealth](conv1) -- node {} (conv2);
\draw [->,>=stealth](conv2) -- node {} (conv3);
\draw [->,>=stealth](conv3) -- node {} (ds);

\draw [->,>=stealth](ds) -- node {} (dots);

\draw [->,>=stealth](dots) -- node {} (conv4);
\draw [->,>=stealth](conv4) -- node {} (conv5);
\draw [->,>=stealth](conv5) -- node {} (ds2);

\draw [->,>=stealth](ds2) -- node {} (mb);
\draw [->,>=stealth](mb) -- node {} (conv6);
\draw [->,>=stealth](conv6) -- node {} (lin1);
\draw [->,>=stealth](lin1) -- node {} (lin2);

\draw [->,>=stealth](digit) -- node {} (emb);
\draw [->,>=stealth](emb) |- node {} (conv2);
\draw [->,>=stealth](emb) |- node {} (conv4);

\end{tikzpicture}
\caption{The proposed GAN architecture for limited-length audio generation.
Like in the original StyleGAN, A is a learned affine transformation and B is
a learned per-channel scaling factor, AdaIn is an adaptive instance normalization layer.}
\label{fig:ourgan}
\end{figure}

\begin{figure}[t]
\centering
\includegraphics[height=42mm,trim={45mm 5mm 45mm 10mm},clip]{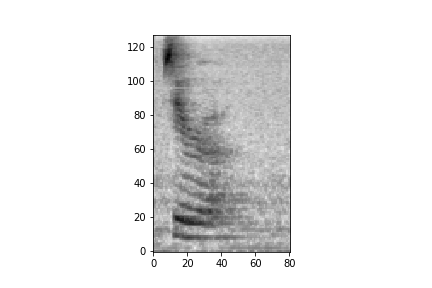}
\includegraphics[height=43mm,trim={30mm 5mm 30mm 8mm}, clip]{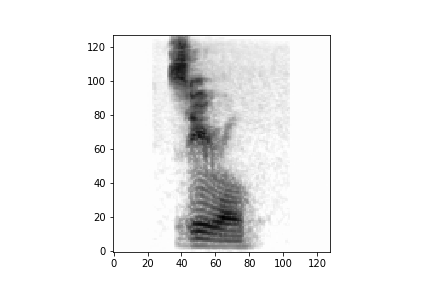}
\\
\includegraphics[height=42mm,trim={45mm 5mm 45mm 10mm},clip]{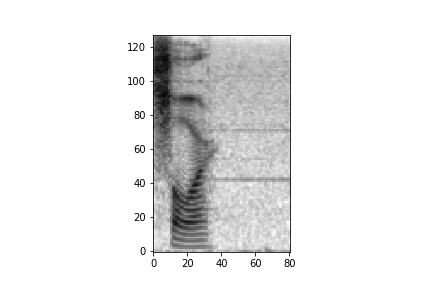}
\includegraphics[height=43mm,trim={30mm 5mm 30mm 8mm}, clip]{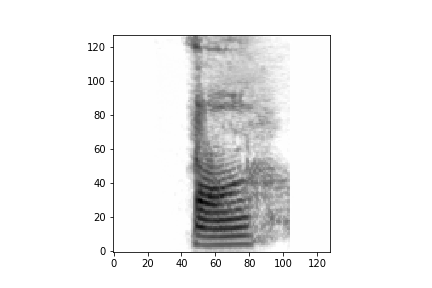}

\caption{Mel-spectrogram of real utterances (left) and mel-spectrograms
generated conditionally on the word (right).
The first row contains examples of word ``zero'' and the second row
contains examples of word ``four''.}
\label{fig:melspec}
\end{figure}

We train a StyleGAN model to generate mel-spectrograms \cite{mels},
which are commonly used to represent audio signals (see Fig.~\ref{fig:melspec} for illustration). Two variants of the model are presented: first, without any labeled conditioning, and second, a conditional model which receives the spoken utterance contents as an additional input.
Our GAN model follows the design of the original StyleGAN with a few differences. The generation starts by sampling a random latent variable 
$
\vz \sim N(0, \matr{I})
$ from the normal distribution. 
Optionally, $\vz$ is appended with a known conditioning vector $\vc$  and transformed with a multilayer perceptron network (a mapping network $f$) to produce a latent code
\[
  \vw = f(\vz, \vc)
\,.
\]
The additional input of the mapping network is the learned embedding of the word (digit class in this paper) that needs to be
generated. The mapping network consists of an eight-layer fully connected network with the leaky ReLu activation
functions and latent code normalization (normalization is done by dividing the latent code by the standard deviation computed from its elements).
Vector $\vc$ is simply used as an extra input to each of the hidden layers of the mapping network \cite{cgan} (see Fig.~\ref{fig:ourgan}).

The latent code $\vw$ is used to modulate the generation process done by the
synthesis network which is a convolutional network that transforms a constant $4 \times 4$ map with $128$ channels into a $128 \times 128$ mel-spectrogram.
Similarly to the original StyleGAN, the synthesis network consists of several convolutional blocks that now operate at different time-frequency scales.
Each block contains an upsampling layer, two convolutional $3 \times 3$ layers, and two adaptive instance normalization layers:
\begin{equation}
  \adain(\vx_i, \vy) = \vy_{s,i} \frac{\vx_i - \mu({\vx_i)}}{ \sigma(\vx_i) } + \vy_{b,i} ,
\label{equation:adain}
\end{equation}
where $\vx_i$ is the $i$-th feature map of the input, $\mu({\vx_i)}$ and $\sigma(\vx_i)$ are the mean and standard deviation computed from $\vx_i$, and $\vy_{s,i}$, $\vy_{b,i}$ are the inputs controlling the normalization process.
The \textit{style} vectors $\vy = (\vy_s, \vy_b)$ are computed using learned affine transformations of the latent codes $\vw$ (blocks A in the figure). To generate extra stochastic details, each block of the synthesis network
contains an independent per-layer noise source, which is a single-channel Gaussian noise image
broadcasted to match all the channels using learned scaling factors (blocks B in Fig.~\ref{fig:ourgan}). 
In the generation process during training, we also use a style-mixing regularization proposed in \cite{stylegan}: blocks in the synthesis network derives their style vector $\vy$ from two different realization of $\vz$. 


Our discriminator is formed of several repetitive blocks as well. Each block contains two convolutional layers with $3\times 3$ kernels and a downsampling layer. The final block in the discriminator starts with a mini-batch standard deviation layer which is followed by a convolution layer and two linear layers. The first linear layer has the leaky ReLu activation function and the second one is plain linear.

We enhance the discriminator by providing it with the information about the desired class of the generated mel-spectrogram. We do this by concatenating the learned embedding of the desired class with intermediate feature maps produced at the beginning of each discriminator block (see Figure~\ref{fig:ourgan}).
%
The upsampling and downsampling layers are implemented as it was done in the original StyleGAN model.
All convolution layers use leaky ReLu activations with a leak factor 0.2 both in the discriminator and the generator.

\section{Experimental setup}

\subsection{Speech commands dataset}

To train the model, we used the Speech Commands dataset \cite{sc09}. The dataset consists of 105,829 utterances of 35 short common words as a one-second or less WAVE format files. The sound samples have been uttered by a large variety of different speakers in different acoustic environments, and the data typically includes only a few samples per speaker.
All samples in the dataset have been quality controlled by rejecting a sample if a human listener could not tell what word was being spoken, or it sounded like an incorrect word.
We decided to use a subset of the dataset to limit the scope of the problem and focused on learning the digits from zero to nine (18,620 samples). The dataset uses a 16\,kHz sample rate. 

\subsection{Conversion between audio and mel-spectrograms}

To compute mel-spectrograms from raw audio, we first generate the linear-frequency spectrogram with the short-time Fourier transform (STFT), then apply a mel-filterbank transformation to map the magnitudes to a mel scale, and finally convert the resulting mel-spectrogram to a decibel scale via a logarithm.  
%
For the STFT, we use  a 50\,ms frame size, 12.5\,ms frame hop, and the Hann window function. The mel filterbank consists of 128 filters spanning from 125\,Hz to 7.6\,kHz, and 
the filterbank output magnitudes are clipped to a minimum value of 0.01 in order to limit the dynamic range in the logarithmic domain and then compressed to log dynamic range. 


For waveform generation from the mel-spectrogram, we first invert the logarithm and apply a mel filterbank pseudoinverse transform to obtain a linear scale magnitude spectrogram, followed by the Griffin-Lim \cite{griffinlim} algorithm to generate the sound samples. Although neural waveform generation methods can result in higher synthesis quality, we opt to use this well-known signal processing method to focus on the mel-spectrogram synthesis performance and exclude potential interaction effects in a fully neural pipeline.

\subsection{Details of model training}
\label{section:modeltraining}

The audio files were first transformed into mel-spectrograms to create the training data for the StyleGAN.
The StyleGAN was trained with many design choices borrowed from \cite{stylegan}.
We used progressive growing to start generating mel-spectrograms at resolution $8 \times 8$ and gradually growing the size of the generated mel-spectrograms to $128 \times 128$ \cite{progan}. The lower-resolution mel-spectrograms were generated by bilinear interpolation of the 128x128 size melspectrograms. The mini-batch size was decreased during the progressive growing to maintain an approximately constant tensor volume in network activations throughout the training. We start with the mini-batch size of 256 for $8 \times 8$ resolution and halve the mini-batch size after we introduce a new higher-resolution layer until we reach the mini-batch size of 32 for the final $128 \times 128$ resolution. Due to the variable batch size, training progress is measured in samples-introduced-to-the-network instead of epochs or minibatch iterations. During progressive growing, we introduce 200,000 mel-spectrograms while fading in a new layer, and train the network for another 200,000 samples after the layer has been fully faded in. After progressive growing is finished, we train the network until 4.05M mel-spectrograms have been introduced to the network in total. This takes about three days on one NVIDIA Tesla V100 GPU. 

We trained the StyleGAN using the WGAN-GP loss \cite{wgangp} with  various modifications proposed in \cite{progan}. 
The discriminator and generator were optimized using minibatch updates at equal update schedules (i.e., the discriminator is updated once for each update of the generator).
%
Additionally, we included a regularization loss term on the mean-squared discriminator activations on the real data to prevent the training from engaging in a ``magnitude race'' described in \cite{progan}.
Specifically, we used an extra term $\epsilon E_{x \in \text{real}}[D(x)^2]$ in the discriminator loss which prevents a magnitude drift of the discriminator output $D(x)$ ($\epsilon$ was set to 0.001).

All weights in convolutional, fully-connected, and affine transform layers were initialized with values drawn randomly from the standard normal distribution. All bias terms and the constant $4 \times 4$ feature map in the synthesis network were initialized with zeros.
We used the Adam optimizer \cite{adam} with $\alpha$ = 0.001, $\beta_1$ = 0.0, $\beta_2$ = 0.99, and $\epsilon$ = $10^{-8}$. However, for the final $128 \times 128$ resolution, we increased the learning rate to 0.0015. Following \cite{stylegan}, we also reduced the learning rate for the mapping net by two orders of magnitude.
We did not use an exponential running average for the weights of the generator.

\section{Results}

We evaluate the quality of the generated audio samples using two objective metrics: the first method 
builds upon the Fr\'{e}chet inception distance \cite{fid} and the second method evaluates the
error rates of the output of the Deep Speech automatic speech recognition system \cite{deepspeech_paper, deepspeech_repo} run on the generated audio samples. Furthermore, we conducted a MOS listening test to evaluate a the subjective quality of the generated samples.

\subsection{Evaluation with Fr\'{e}chet distance}
\label{section:fid}

Fr\'{e}chet distance (FD) \cite{fid} is commonly used in GAN research to measure the quality and variation of generated samples by examining the activation statistics of a pre-trained classifier model. In image generation applications, an Inception net classifier is typically used \cite{szegedy2016rethinking}, but we found that simply viewing spectrograms as monochrome images provided inconsistent results. Instead, we use two domain-specific classifiers to score the generated samples in terms of content and speaker variation.

To score the generated samples in terms of content, we trained a classifier on the spoken digits dataset to recognize the ten different digit classes from mel-spectrograms. The architecture of the classifier was similar to the one used in \cite{wavegan} to classify melspectrograms to compute the inception score. We modified that architecture by adding an extra average-pooling layer to produce the activations used for computing the Fr\'{e}chet distance (FD) score. The classifier achieved a 97\% accuracy on the test set of 2552 samples after training with 150,000 mel- spectrograms.
The FD score was computed as
\begin{equation}
\text{FD} = {\lVert \vm_r-\vm_g \rVert}_{2}^2 + \text{Tr}(\mC_r + \mC_g - 2(\mC_r\mC_g)^{1/2}),
\label{equation:fid}
\end{equation}
where $\text{Tr}()$ denotes the trace of a matrix,
$\vm_r$, $\mC_r$ are the mean and covariance matrix computed from the classifier activations on real data samples, and $\vm_g$, $\mC_g$ are the same statistics computed from the classifier activations on the generated data.




Table~\ref{table:fid} presents FD scores computed using the described digit-classifier for different designs of the StyleGAN generator.
The scores were calculated multiple times during the course of training using the training set (18,620 mel-spectrograms). The lowest values of the FD scores are reported in Table~\ref{table:fid}.
We can clearly see that progressive growing and style-mixing regularization improve the FD score. Label conditioning slightly decreases the quality of the generated mel-spectrograms but gives a way to control the generation process.

To score the generated samples in terms of speaker variation, we computed the FD score \eqref{equation:fid} using the activations of a pre-trained speaker embedding model from  \cite{kaseva2019-spherediar}. The motivation for using this kind of embeddings was to provide means for measuring variation of speaker identity in the generated samples regardless of the contents (conditioning input $\vc$) of the utterances. This speaker information was unlabeled in the present experiments, but the generative model should ideally learn to embed such variability in the latent code $\vz$. 
The results are presented in Table~\ref{tab:speaker_embedding_fd}. We can see that the StyleGAN model outperforms the comparison WaveGAN \cite{wavegan} method.

\begin{table}[t]
\caption{Fr\'{e}chet distance for various generator designs (the smaller the better).}
\label{table:fid}
\centering
\begin{tabular}{ccccc}
\hline
Model & Label & Progressive & Style-mix & \textbf{FD}  \\ 
& cond. & growing & regul. & \\ 
\hline 
StyleGAN-U1 & no & no & no & $49.0$  \\ 
StyleGAN-U2 & no & yes & no & $\mathbf{27.1}$  \\
\hline
StyleGAN-C1 & yes & yes & no & $41.6$ \\ 
StyleGAN-C2 & yes & yes & yes & $\mathbf{31.3}$  \\ 
\hline 
\end{tabular}
\end{table}

\begin{table}[t]
\caption{Fr\'{e}chet distance in speaker embedding space (the smaller the better).}
\label{tab:speaker_embedding_fd}
\centering
\begin{tabular}{lc}
\hline
\textbf{Method} & \textbf{FD}  \\   
\hline 
Griffin-Lim reconstructed &  0.11 \\
StyleGAN (proposed) & 0.24 \\
WaveGAN & 0.33 \\
\end{tabular}
\end{table}

\subsection{Evaluation with the Deep Speech recognizer}

FD is a measure which assesses the quality of unconditional generation of  mel-spectrograms. To evaluate the quality of the conditional generation, we convert the generated mel-spectrograms into audio with the Griffin-Lim transform and then attempt to decode the audio to text using a pre-trained Deep Speech end-to-end speech recognition system \cite{deepspeech_paper, deepspeech_repo}. Thus, the Deep Speech recognizer performance acts as a proxy for intelligibility evaluation. In this case, the metric for conditional generation quality is the character error rate (CER)~\cite{cer} between the desired sequence of characters and the sequence produced by the Deep Speech recognizer.

\begin{figure}[t]
\centering
Total CER scores\\
\includegraphics[width=\linewidth,trim={6mm 7mm 15mm 10mm},clip]{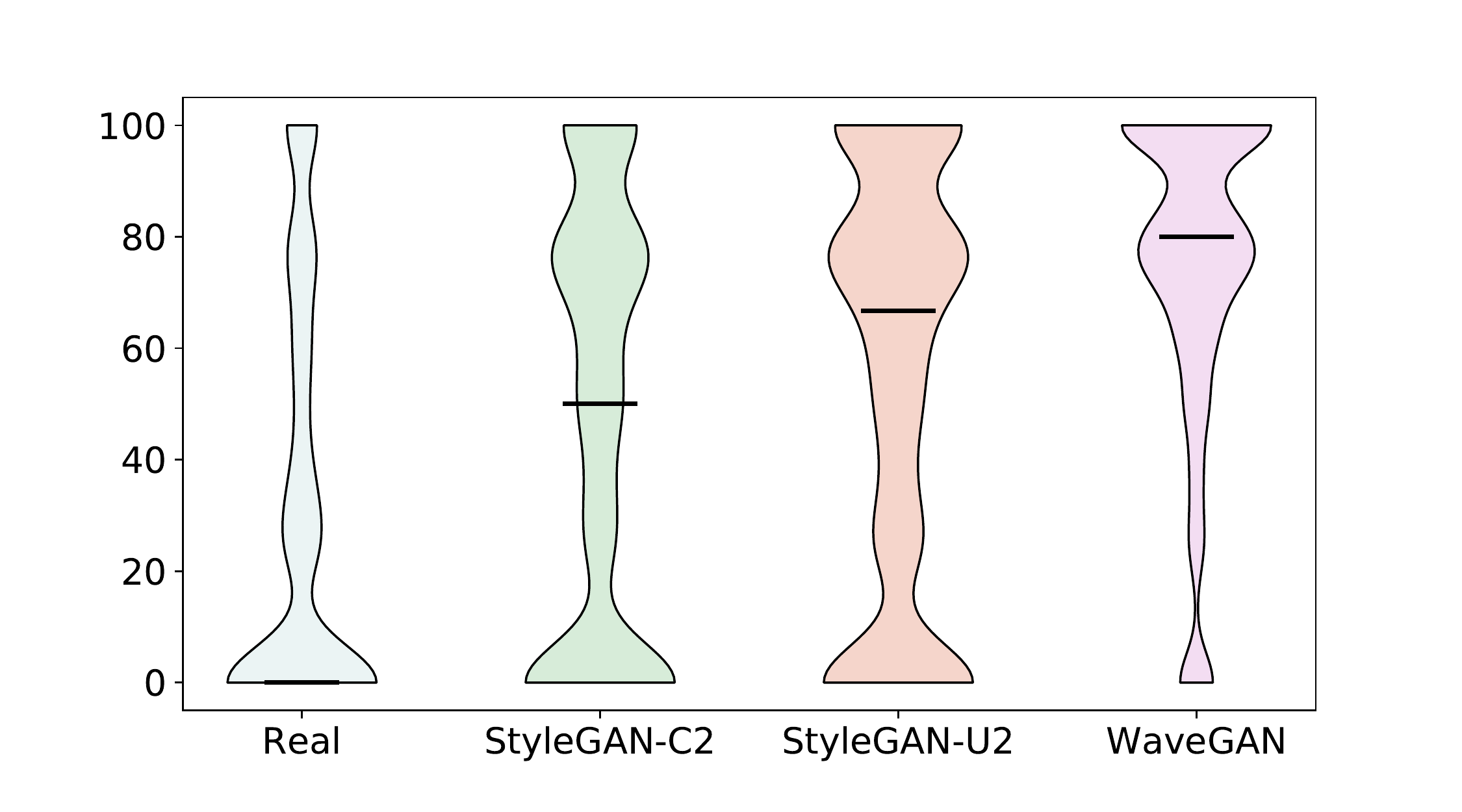}
\caption{The violin plot of the total CER scores of the (generated) audio samples converted into text with the Deep Speech recognizer (the smaller the better). The thick black line represents the median CER scores.}
\label{fig:cer_total}
\end{figure}

We generated 500 samples for each digit with each assessed generator design and label the generated samples with 10 digit classes utilizing the classifier introduced in Section~\ref{section:fid}. For conditionally-generated samples we also used labels produced by the classifier instead of the ones used in the generation process. 
As the baseline, we also computed the CER scores for random samples from the training data. In addition, we computed the scores for the samples generated by the current state-of-the-art GAN in waveform generation, the WaveGAN \cite{wavegan}, using a pre-trained model provided by the authors \cite{wavegan_repo}.

The results are presented in Fig.~\ref{fig:cer_total}. We can see that StyleGAN models achieved lower CER scores compared to the WaveGAN. In addition, the StyleGAN model with label conditioning outperformed the one without conditioning.


\subsection{Listening test}

For subjective quality evaluation, we conducted a listening test on the Amazon Mechanical Turk crowdsourcing platform (limiting workers by location to the English speaking countries). 
The listeners were presented with samples from each system under evaluation and asked to rate the naturalness of the sample on a five-point absolute category rating scale ranging from 1 (Bad) to 5 (Excellent). Four systems were included in the comparison: ``Natural'' samples are unprocessed utterances from the dataset, while ``Re-synthesis'' samples are synthesized from unmodified mel-spectrograms of natural samples. This represents the upper limit in quality using the present waveform synthesis method.
%

\begin{figure}[t]
\centering
\includegraphics[width=\linewidth]{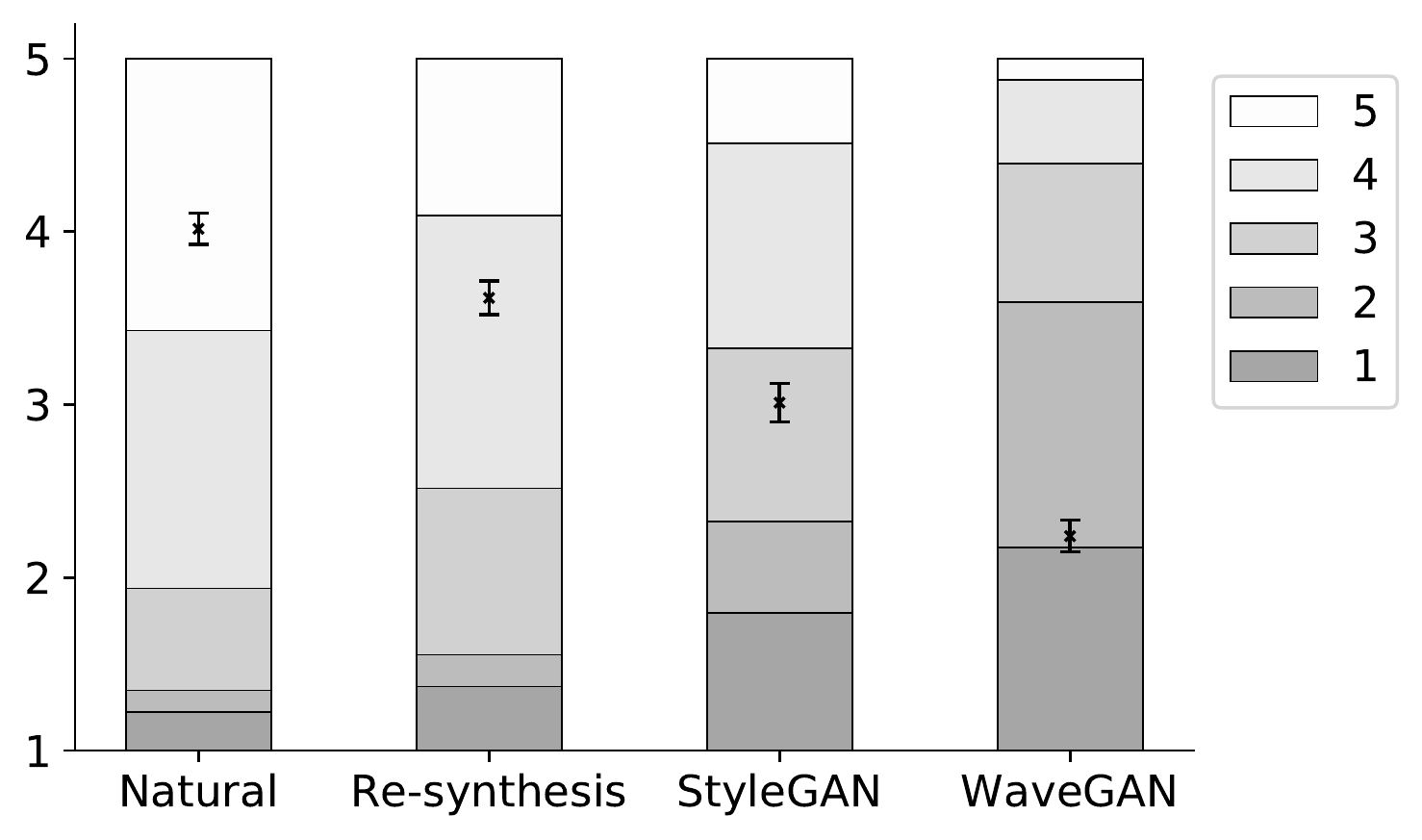}
\caption{Naturalness mean opinion score (MOS) ratings with 95\% confidence intervals. Stacked distribution histograms for the ratings are shown in the background.}
\label{fig:mos-test}
\end{figure}

The tests were split into eight batches of 100 test cases, and each batch was evaluated by five individual workers. A total number of 3946 valid evaluations was collected in the listening test. Samples were drawn randomly for each system, but  balanced between different digit classes. 
Figure \ref{fig:mos-test} shows mean opinion scores (MOS) for naturalness with t-statistic based 95\% confidence intervals Bonferroni corrected for multiple comparisons. Stacked histograms for the answer distributions are shown in the background. 
In the plot, the ratings have been averaged over listeners and digits. 
The results show that the proposed method StyleGAN-C2 outperforms WaveGAN.  

\section{Conclusion}

In this  paper, we adapted the recently developed StyleGAN  model for speech generation with minimal or no conditioning on text. The proposed model produced higher-quality audio samples and captured better the data distribution compared to previous GAN-based speech generation models.
%
There are two clear obstacles for the use of the presented model for generic text-to-speech mapping, but these are left as future work. First, the present model uses static conditioning, whereas a TTS system input is typically a sequence of characters (or phonemes). A natural extension would be to include a sequence encoder similar to Tacotron \cite{tacotron2}, perhaps combined with a convolutional self-attention mechanism \cite{zhang2019-sagan}.
Second, the synthesis network currently generates a fixed-length spectrogram output. Introducing a duration prediction model to generate the synthesis network input feature map could allow varying output lengths, as the network itself is otherwise fully convolutional.
Another interesting direction is to move towards end-to-end training and eliminate the intermediate step of conversion between raw audio and mel-spectograms, which may potentially further improve the quality of the generated samples.

Source code and audio samples are available at  \url{https://github.com/kapalk/cStyleGAN}

\section{Acknowledgements}
We acknowledge the computational resources provided by the Aalto Science-IT project.

\bibliographystyle{IEEEtran}

\bibliography{sources}

\end{document}